\documentclass[sigconf]{acmart} 

\renewcommand\footnotetextcopyrightpermission[1]{} 
\AtBeginDocument{%
  \providecommand\BibTeX{{%
    \normalfont B\kern-0.5em{\scshape i\kern-0.25em b}\kern-0.8em\TeX}}}

\usepackage{multirow,makecell}
\usepackage{tcolorbox}
\usepackage{color,xcolor}
\usepackage{listings,amsfonts}
\usepackage{caption}
\usepackage{subcaption}
\usepackage{threeparttable}
\usepackage{bbding}
\usepackage{graphicx}
\usepackage{booktabs} 
\usepackage{longtable}
\usepackage{enumitem}
\usepackage{framed}
\definecolor{shadecolor}{rgb}{0.98,0.95,0.96}
\usepackage{tikz}
\usepackage{pifont}
\usepackage{listings}
\newcommand{\cmark}{\ding{51}}

\newcommand{\yes}{\raisebox{-0.15em}{\tikz\draw[fill=black] (0,0) circle (.38em);}}
\newcommand{\no}{\raisebox{-0.15em}{\tikz\draw[fill=white] (0,0) circle (.38em);}}

\definecolor{customcite}{HTML}{b83b5e}
\definecolor{customlink}{HTML}{07689f}
\definecolor{customurl}{HTML}{11999e}
\AtEndPreamble{
\usepackage{hyperref}

    \hypersetup{
      colorlinks = true,
      linkcolor = customlink,
      anchorcolor = purple,
      citecolor = customcite,
      filecolor = purple,
      urlcolor = customurl
    }
}

\usepackage{tcolorbox}
\def\BibTeX{{\rm B\kern-.05em{\sc i\kern-.025em b}\kern-.08em
    T\kern-.1667em\lower.7ex\hbox{E}\kern-.125emX}}

\newcommand{\tool}{\textsc{MCP-BiFlow}}

\begin{document}

\title{Unsafe by Flow: Uncovering Bidirectional Data-Flow Risks in MCP Ecosystem}


\author[X Hou]{Xinyi Hou}
\email{xinyihou@hust.edu.cn}
\affiliation{%
  \institution{Huazhong University of Science and Technology}
  \city{Wuhan}           
  \country{China}
}

\author[Y Zhao]{Yanjie Zhao}
\email{yanjie_zhao@hust.edu.cn}
\affiliation{%
  \institution{Huazhong University of Science and Technology}
  \city{Wuhan}
  \country{China}
}

\author[H Wang]{Haoyu Wang}
\email{haoyuwang@hust.edu.cn}
\authornote{Haoyu Wang is the corresponding author (haoyuwang@hust.edu.cn).}
\affiliation{%
  \institution{Huazhong University of Science and Technology}
  \city{Wuhan}           
  \country{China}
}

\begin{abstract}
Model Context Protocol (MCP) have quickly become the interface layer between LLM agents and external tools, yet they also introduce unsafe data flows that existing analyzers handle poorly. Vulnerabilities manifest in two directions: requester-controlled arguments may propagate to sensitive operations, while untrusted external or sensitive internal data may surface through MCP-visible outputs and subsequently influence host or model behavior. Accurate detection is complicated by the heterogeneous registration and dispatch patterns MCP servers employ, the need for MCP-specific taint semantics, and the fact that bugs often only materialize along complete tool-scoped execution paths. We present \tool{}, a bidirectional static analysis framework built around MCP-aware entrypoint recovery, protocol-specific taint modeling, and interprocedural propagation analysis. Against a benchmark of 32 confirmed MCP vulnerability cases, \tool{} identifies 30 (93.8\% recall), substantially outperforming CodeQL, Semgrep, Snyk Code, and MCPScan. Across 15,452 real-world MCP server repositories, \tool{} surfaces 549 overlap-compressed candidate clusters; manual review confirms 118 vulnerability paths in 87 servers, establishing unsafe propagation as a recurring failure mode that resists detection without protocol-aware recovery of both request-side and return-side flows.
\end{abstract}

\maketitle

\section{Introduction}

Large language model (LLM) applications are rapidly evolving from traditional chatbots into agents that can use tools and interact with external environments, e.g., for web retrieval, filesystem access, database querying, and code execution~\cite{li2024personal,xi2023risepotentiallargelanguage,gan2024navigating,wang2025largemodelbasedagents,weng2023agent}. This shift is evident in agent frameworks, e.g., LangChain~\cite{LangChain}, AutoGen~\cite{wu2023autogen}, and CrewAI~\cite{crewai2025}, in coding agents, e.g., Claude Code~\cite{anthropic2026claudecode} and OpenAI Codex~\cite{openai2025codex}, and in general-purpose assistants, e.g., Manus~\cite{manus_2025} and OpenClaw~\cite{openclaw_2025}. However, incompatible tool interfaces, schemas, and orchestration logic across platforms and frameworks hinder the portability and reuse of LLM agents~\cite{hou2025mcp,openaiplugin,cozeplugin,LangChain}. To address this issue, Anthropic introduced the Model Context Protocol (MCP) in late 2024~\cite{anthropic_mcp_2024}, a standardized host--client--server architecture in which servers expose tools, resources, and prompts to LLM applications through a uniform interface~\cite{hou2025mcp}. Within 16 months, the number of MCP servers has grown to nearly 60k~\cite{mcpworld}. Their utility is evidenced by strong adoption, with Microsoft's Playwright MCP server exceeding 29k GitHub stars~\cite{playwright_mcp_2025} and the community-maintained Figma MCP server reaching about 14k~\cite{figma_context_mcp_2025}.

The growing adoption of MCP servers also has important security implications. In the MCP workflow, user input is received by an MCP host, where it is processed by an agent that decides whether to invoke a tool through an MCP client. The client then sends invocation arguments to the MCP server, which may access privileged local or remote resources and return results to the client, the host, or the end user. This workflow crosses multiple trust boundaries, as data flows from requester-controlled inputs and external content sources to privileged operations and downstream use of tool outputs.
We therefore view MCP server security as a bidirectional trust-boundary data-flow problem. The first direction is \textbf{request-side propagation}, in which requester-controlled data is incorporated into tool parameters such as URLs, file paths, shell arguments, or query strings before reaching sensitive operations. Such flows can lead to vulnerabilities including SSRF, path traversal, command injection, and unsafe query execution. CVE-2025-53107 in \texttt{git-mcp-server}, for example, shows that git-related inputs can reach \texttt{child\_process.exec}~\cite{nvd-cve-2025-53107}. The second direction is \textbf{return-side propagation}, in which untrusted external content or sensitive internal data is incorporated into tool outputs and then consumed by the client, the host, or the model. CVE-2025-53355 in \texttt{mcp-server-kubernetes}, for example, shows that pod log content returned by the server can later influence a tool invocation~\cite{nvd-cve-2025-53355}. In MCP settings, both directions are security-relevant because tool inputs may trigger privileged effects and tool outputs may re-enter LLM-mediated reasoning.

Existing work does not fully address this setting. Recent MCP security studies have mainly characterized ecosystem risks, attack patterns, and defenses~\cite{hou2025mcp,hasan2025mcpglance,zhao2025mcpattack,radosevich2025mcpsafety}, while general-purpose analyzers such as CodeQL~\cite{codeql} and Semgrep~\cite{semgrep} are not designed to recover MCP-specific interaction structure end to end. Systematic MCP analysis faces three concrete difficulties. (1) MCP servers expose externally reachable functionality through diverse tool registration, protocol-dispatch, and wrapper patterns rather than stable route abstractions. (2) Generic taint templates are often mismatched to MCP trust boundaries: on the request side, the relevant source is typically the decoded MCP argument payload, whereas on the return side the source may be externally fetched content and the sink is the protocol-visible return crossing the MCP boundary. (3) MCP vulnerabilities rarely manifest as short local source-to-sink pairs; they emerge through complete tool-scoped interaction paths spanning argument decoding, validation, helper forwarding, sensitive operations, and protocol-visible outputs.

We address these challenges with \tool{}, a bidirectional static analysis framework for MCP servers. \tool{} begins by recovering tool-specific entrypoints, linking publication sites with the dispatch logic that routes a given invocation, so analysis is anchored to the handler reachable for each tool rather than a generic protocol wrapper. It then assigns taint semantics matched to MCP trust boundaries, distinguishing decoded request arguments, externally obtained values, sensitive operations, and protocol-visible returns. Finally, bidirectional interprocedural analysis over these recovered boundaries reconstructs full request-side and return-side propagation paths. The pipeline remains largely deterministic: LLM assistance is used only when local syntax is insufficient to determine whether a source remains materially attacker-controlled or whether a guard meaningfully blocks the flow.
We evaluate \tool{} on both curated vulnerability cases and a broad corpus of open-source MCP servers. On a benchmark of 32 confirmed MCP vulnerability cases, \tool{} detects 30 (93.8\% recall), substantially outperforming CodeQL, Semgrep, Snyk Code, and MCPScan. Across 15,452 real-world MCP server repositories, the tool produces 549 overlap-compressed candidate clusters spanning 424 servers; disclosure-oriented manual review confirms 118 vulnerability paths in 87 servers. 
In summary, we make the following contributions:

\begin{itemize}[leftmargin=15pt]
    \item \textbf{Propagation-based view of MCP vulnerabilities.} We frame MCP server security around two directions of unsafe propagation: \emph{request-side propagation} and \emph{return-side propagation}. This view captures how requester-controlled inputs, untrusted external content, and sensitive internal data cross trust boundaries to reach sensitive operations or model-consumable outputs.

    \item \textbf{Practical analysis framework.} We build \tool{}, which detects unsafe propagation in MCP servers by combining MCP-aware entrypoint recovery, protocol-specific taint modeling, and bidirectional interprocedural analysis. On a benchmark of 32 confirmed vulnerabilities, \tool{} reaches 93.8\% recall.

    \item \textbf{Large-scale study of real-world MCP servers.} We apply \tool{} to 15,452 MCP servers across Python, JavaScript, and TypeScript projects. The pipeline surfaces 549 overlap-compressed candidate clusters across 424 servers, and final review confirms 118 vulnerability paths across 87 servers.
\end{itemize}

\noindent\textbf{Artifact Availability.} \tool{} and all artifacts are available at \url{https://doi.org/10.5281/zenodo.19335837}.

\section{Background and Motivation}

\subsection{LLM Powered Autonomous Agents}

LLM-powered autonomous agents have evolved from prompt-driven reasoning systems into software that can plan, retain state, and act on external environments~\cite{weng2023agent,yao2023react}. A typical architecture centers on an LLM augmented with planning, memory, and action modules, and among these components, \emph{tool use} has emerged as the primary mechanism for capability expansion. By invoking APIs, querying databases, executing code, browsing the web, or interacting with desktop environments, agents can operate on real software and real-world resources rather than text alone~\cite{schick2023toolformer,langgraph2024,anthropic2024computeruse,openai2025functioncalling}.
The scale of this shift is visible across the ecosystem. Frameworks such as LangChain and LangGraph have lowered the barrier to building long-running and multi-agent workflows~\cite{langchain2024v01,langgraph2024}; coding agents like Claude Code and Codex have brought tool use into real repositories and developer environments~\cite{anthropic2026claudecode,openai2025codex,openai2026codexapp}; and general-purpose platforms such as OpenClaw have begun integrating chat interfaces, local execution, connected services, and reusable skills into unified agent ecosystems~\cite{openclaw2026intro,openclaw2026trust}. Alongside this growth, the mechanism of tool use has itself changed, moving from application-specific function calling and framework-level abstractions toward more reusable and interoperable interfaces, including protocolized connectors such as MCP and modular capability packages such as skills~\cite{openai2025functioncalling,hou2025mcp,openai2026skills,openai2026codexskills}.


\subsection{The Rise of Model Context Protocol}

\begin{figure}[t]
    \centering
    \includegraphics[width=1\linewidth]{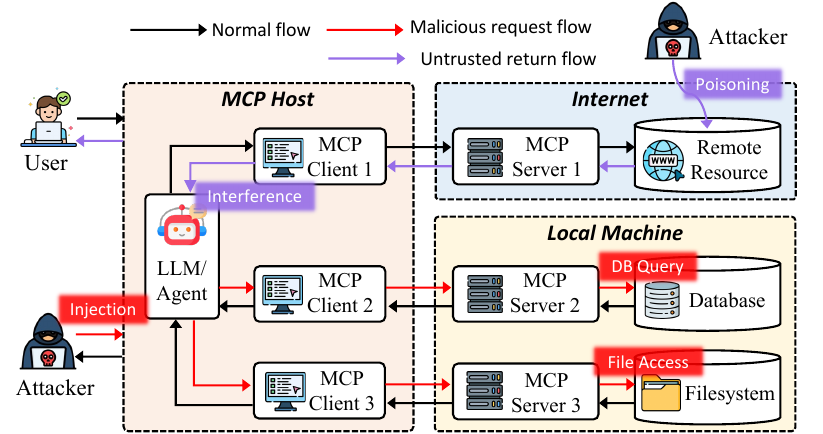}
    \caption{Data flows in MCP systems.}
    \label{fig:background}
\end{figure}

Model Context Protocol (MCP), introduced by Anthropic in late 2024, is an open standard for connecting AI applications to external tools, data sources, and workflows through a unified protocol interface~\cite{anthropic2024mcp,mcpdocs2025intro}. Compared with earlier function-calling APIs and framework-specific tool abstractions, MCP shifts tool integration toward protocol-level interoperability, including capability discovery, negotiation, and reuse across heterogeneous systems~\cite{anthropic2024mcp,mcpdocs2025architecture,hou2025mcp}. Since then, the ecosystem has expanded from mostly local integrations to standardized SDKs, broader platform adoption, and remote deployments~\cite{anthropic2025codeexecution,mcpblog2025transportfuture,mcpdocs2025spec}.
At a high level, MCP involves three roles: an \emph{MCP host}, an \emph{MCP client}, and an \emph{MCP server}. The host is the user-facing AI application, the client is the protocol component inside the host, and the server exposes tools, resources, or prompts~\cite{mcpdocs2025clients,mcpdocs2025architecture,hou2025mcp}. In a typical workflow, as shown in \autoref{fig:background}, the host decides whether an external capability is needed, the client discovers and invokes the corresponding server-side functionality, and the result is returned for further reasoning or final response generation~\cite{mcpdocs2025clients,mcpdocs2025architecture,mcpdocs2025lifecycle,hou2025mcp}. 
This architecture improves extensibility, but it also creates new failure modes: user-controlled inputs may reach privileged server-side operations, while untrusted external content or sensitive internal data may flow back through MCP outputs and affect downstream reasoning or user-visible responses~\cite{nvd-cve-2025-5277,ghsa-cve-2025-53110,ghsa-cve-2025-53107,ghsa-cve-2025-53355}.

\begin{table*}[t]
\centering
\fontsize{6.9}{8.3}\selectfont
\caption{Main tool registration and dispatch patterns in MCP servers.}
\label{tab:entrypoint_patterns}

\renewcommand{\arraystretch}{1.10}
\setlength{\tabcolsep}{3.0pt}

\begin{tabular}{|c|p{5.4cm}|c|p{5.2cm}|c|}
\hline
\multicolumn{1}{|c|}{\textbf{Pattern}} &
\multicolumn{1}{c|}{\textbf{Python Example}} &
\multicolumn{1}{c|}{\textbf{\# Python}} &
\multicolumn{1}{c|}{\textbf{JS/TS Example}} &
\multicolumn{1}{c|}{\textbf{\# JS/TS}} \\
\hline
\textbf{Direct declaration}
& \begin{tabular}[t]{@{}l@{}}
\texttt{@<server\_instance>.tool(name="x")} \\
\texttt{def tool(...): ...}
\end{tabular}
& 54
& \begin{tabular}[t]{@{}l@{}}
\texttt{<server\_instance>.tool("x", desc, schema,} \\
\texttt{\ \ async (args) => \{ ... \})}
\end{tabular}
& 30 \\
\hline
\textbf{Explicit registration APIs}
& \begin{tabular}[t]{@{}l@{}}
\texttt{<server\_instance>.add\_tool(tool\_fn, name="x")} \\
\texttt{/* imperative registration */}
\end{tabular}
& 1
& \begin{tabular}[t]{@{}l@{}}
\texttt{<server\_instance>.registerTool("x", \{...\},} \\
\texttt{\ \ async (args) => \{ ... \})}
\end{tabular}
& 14 \\
\hline
\textbf{Protocol-level dispatch}
& \begin{tabular}[t]{@{}l@{}}
\texttt{@<server\_instance>.list\_tools()} \\
\texttt{@<server\_instance>.call\_tool()} \\
\texttt{async def call\_tool(...): ...}
\end{tabular}
& 13
& \begin{tabular}[t]{@{}l@{}}
\texttt{<server\_instance>.setRequestHandler(} \\
\texttt{\ \ ListToolsRequestSchema, ...)} \\
\texttt{<server\_instance>.setRequestHandler(} \\
\texttt{\ \ CallToolRequestSchema, ...)}
\end{tabular}
& 44 \\
\hline
\textbf{Other / mixed}
& \begin{tabular}[t]{@{}l@{}}
\texttt{/* custom abstraction, alternate} \\
\texttt{\ \ implementation, or no matched} \\
\texttt{\ \ tool-entry pattern */}
\end{tabular}
& 32
& \begin{tabular}[t]{@{}l@{}}
\texttt{/* custom wrapper, mixed styles,} \\
\texttt{\ \ or no matched tool-entry} \\
\texttt{\ \ pattern in quick pass */}
\end{tabular}
& 12 \\
\hline
\textbf{Total} & & \textbf{100} & & \textbf{100} \\
\hline
\end{tabular}
\end{table*}

\begin{figure*}[t]
    \centering
    \includegraphics[width=0.93\linewidth]{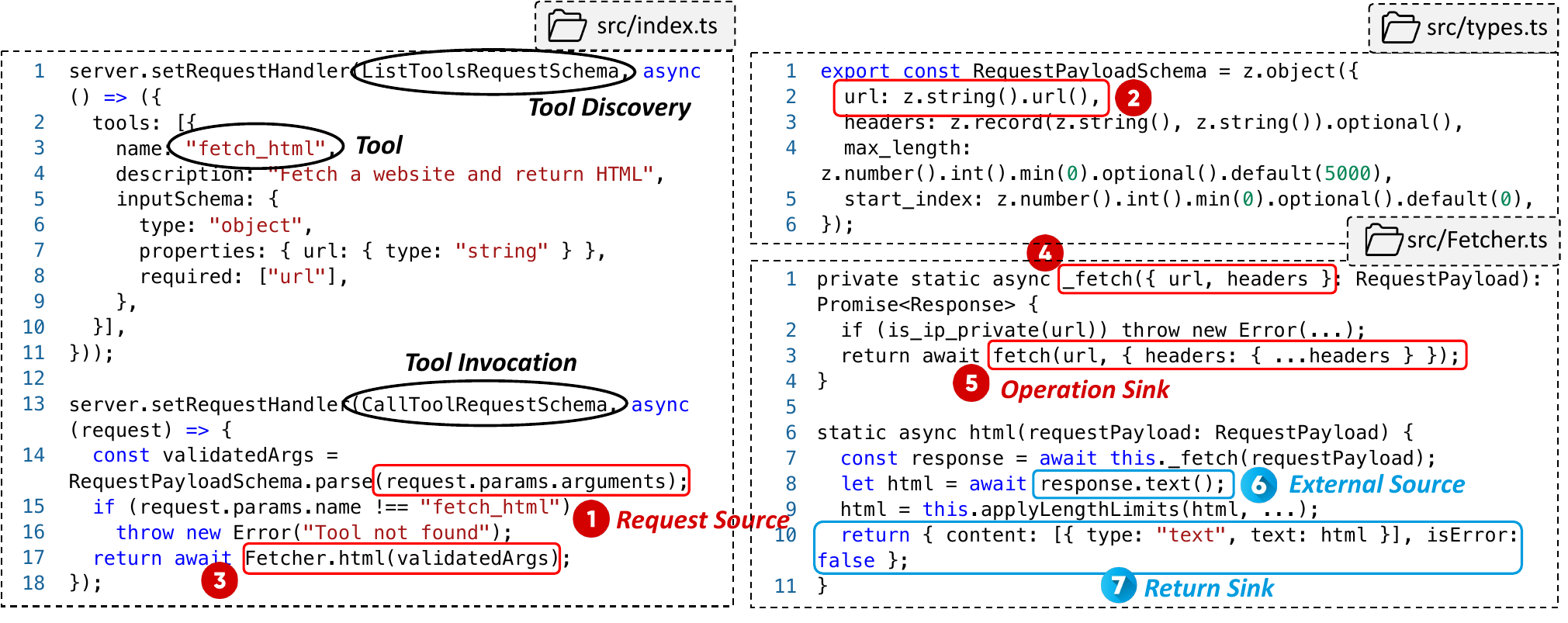}
    \caption{Running example based on CVE-2025-65513, marking request-side (\textcolor{red}{red}) and return-side (\textcolor{blue}{blue}) propagation paths.}
    \label{fig:example}
\end{figure*}

\subsection{Problem Statement}

\noindent\textbf{Problem verview.}
In an MCP workflow, requester-controlled inputs may cross the host--client--server boundary and eventually influence privileged server-side operations, while externally obtained or sensitive server-side data may cross the MCP boundary through tool outputs and later affect the client, the host, the model, or the end user, as illustrated in \autoref{fig:background}. We thus formulate MCP server security as a bidirectional trust-boundary data-flow problem.

\noindent\textbf{Threat model.}
We assume that the MCP host, client, and server are implemented by non-malicious parties and run in a non-compromised environment. Our focus is therefore not malicious infrastructure, but unsafe propagation caused by incomplete validation, insufficient sanitization, or unsafe handling of externally obtained or internally sensitive data. The attacker does not control server internals, source code, the host machine, or the deployment pipeline. Instead, the attacker acts through normal interaction channels, either by providing crafted inputs through the user-facing interface or by controlling external content sources consumed by MCP servers, such as third-party APIs, public repositories, web pages, or remote databases.
Under this model, we consider two classes of unsafe propagation. \textbf{Request-side propagation} occurs when requester-controlled data reaches privileged or security-sensitive operations, including command execution, filesystem access, database queries, code evaluation, and outbound network requests. \textbf{Return-side propagation} occurs when untrusted external content or sensitive internal data is exposed through MCP-visible outputs, thereby crossing the server boundary and becoming available to the client, the host, or the end user. In MCP settings, such return-side flows are security-relevant because they may expose sensitive data or enable mixed chains in which returned content later affects downstream model reasoning or tool invocation. We exclude attacks that depend on a compromised code base, supply chain, or deployment infrastructure, as well as issues of authentication, authorization, and business logic unless they manifest as unsafe cross-boundary data flows under this model.

\begin{figure*}[t]
    \centering
    \includegraphics[width=1\linewidth]{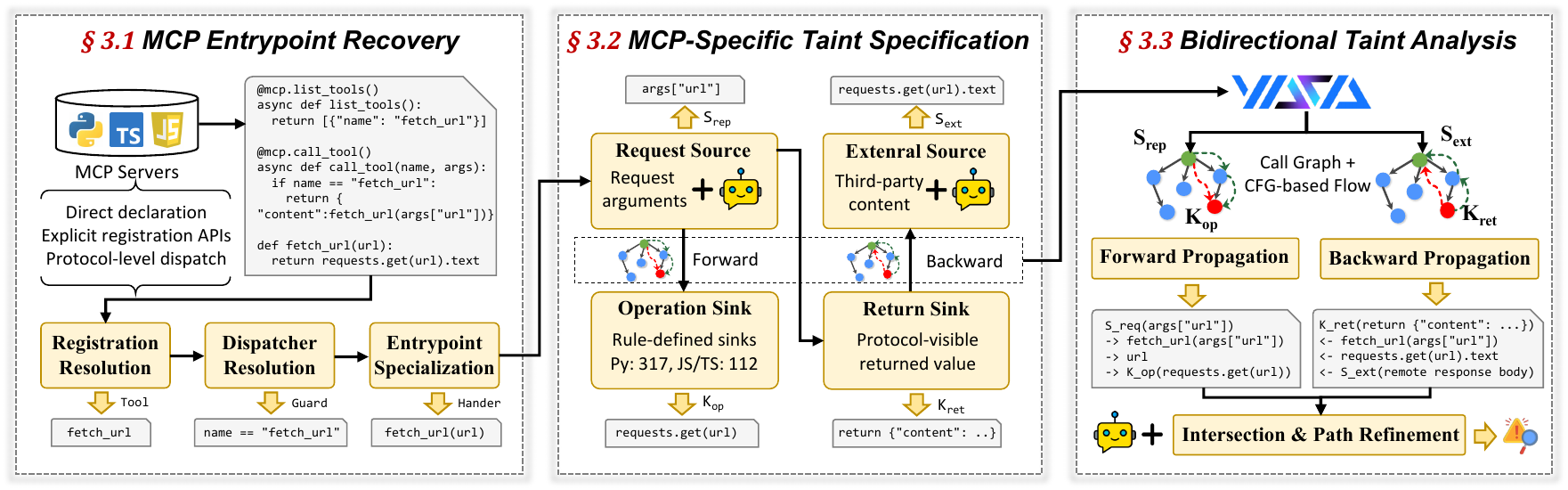}
    \caption{Overview of the \tool{} pipeline.}
    \label{fig:methodology}
\end{figure*}

\subsection{Static Analysis Challenges in MCP Servers}

\noindent\textbf{Entrypoint discovery.}
Static analysis tools usually rely on framework models to identify externally reachable code, such as route decorators, handler registration APIs, and configuration driven bindings. Recovering these entrypoints matters because they mark where external input enters application logic and therefore seed reachability and taint analysis. This works well for frameworks such as Flask and FastAPI, where routes are explicit, but it often misses MCP servers, where externally reachable functionality is exposed through tool registration and dispatch.
This limitation is beginning to receive attention in practice. Snyk Code, for example, has added support for recognizing MCP entrypoints~\cite{snyk-code-support-mcp}, but current support mainly targets simple patterns such as \texttt{FastMCP} declarations. Real-world MCP servers use a much broader range of registration and dispatch styles. To characterize this diversity, we manually analyzed 100 Python and 100 JS/TS projects from the two largest MCP server ecosystems. \autoref{tab:entrypoint_patterns} summarizes representative patterns. In the Python SDK, tools may be exposed through \texttt{FastMCP} decorators or lower level handlers such as \texttt{list\_tools()} and \texttt{call\_tool()}. In the TypeScript SDK, tools may be exposed either through direct registration APIs such as \texttt{tool(...)} or through protocol request handlers such as \texttt{setRequestHandler(CallToolRequestSchema, ...)}. Example in \autoref{fig:example} shows the latter pattern.

\begin{shaded}
\noindent\textbf{Challenge 1:} MCP entrypoints are exposed through diverse registration and dispatch patterns, making them difficult to recognize with existing static analysis models.
\end{shaded}

\noindent\textbf{Taint semantics.}
Recovering tool entrypoints alone is not enough to detect vulnerabilities in MCP, because generic repository-wide taint policies are too coarse for this setting. On the request side, many sensitive sinks remain relevant, but the corresponding sources are often hidden by MCP-specific protocol decoding, branch dispatch, schema parsing, and local aliasing. In \autoref{fig:example}, the relevant source is not the entire \texttt{request} object, but the MCP argument payload in \texttt{request.params.arguments}. On the return side, the sink is defined by the recovered tool boundary, but the source often does not come from client input; it may instead be externally obtained content such as \texttt{response.text()} that is later exposed through the MCP response. Generic taint specifications do not model these request-boundary values and external content sources well.

\begin{shaded}
\noindent\textbf{Challenge 2:} Generic source and sink templates do not capture the MCP-specific taint semantics needed to model request-side and return-side propagation.
\end{shaded}

\noindent\textbf{Tool-scoped path recovery.}
MCP tool logic rarely appears as a short local source-to-sink pattern. On the request side, tool arguments are often parsed, validated, repackaged, and forwarded through helper layers before they reach a sensitive sink. On the return side, protocol-visible outputs are often assembled from external responses, local wrappers, and intermediate transformations before crossing the MCP boundary. In \autoref{fig:example}, the red path (1, 3, 4, 5) captures request-side propagation from the MCP argument source \texttt{request.params.arguments} to the outbound \texttt{fetch} sink, while the blue path (6, 7) captures return-side propagation from the external source \texttt{response.text()} to the protocol-visible return. The schema constraint at (2) further complicates the request-side path by refining the value without clearly removing its security relevance. In practice, MCP vulnerabilities are defined by complete tool-scoped interaction paths rather than isolated local patterns. 

\begin{shaded}
\noindent\textbf{Challenge 3:} MCP vulnerabilities often emerge only along complete tool-scoped interaction paths, so local source or sink reasoning alone is not enough.
\end{shaded}

\section{Design of \tool{}}

\label{sec:methodology}

We design \tool{} as a static taint analysis framework for recovering flows that cross trust boundaries in real-world MCP servers. The overall workflow is shown in \autoref{fig:methodology}. 
\tool{} integrates \emph{MCP Entrypoint Recovery (\autoref{subsec:entrypoint})} to identify procedures reachable through tool interactions, \emph{MCP-Specific Taint Specification (\autoref{subsec:source-sink})} to model data derived from requests, externally obtained content, sensitive operations, and protocol-visible returns, and \emph{Bidirectional Interprocedural Taint Analysis (\autoref{subsec:propagation})} to recover complete vulnerability paths from requests to sensitive operations and from external or internal sources to outputs.

\subsection{MCP Entrypoint Recovery}
\label{subsec:entrypoint}

To address \emph{Challenge~1}, \tool{} performs MCP entrypoint recovery before taint propagation. The goal is to recover logical tool entrypoints even when tool publication, dispatch, and execution are implemented in different procedures. \tool{} first resolves publication patterns that expose tools to clients, then resolves dispatcher logic that maps tool names to concrete handlers, and finally combines the two into tool-specific analysis entrypoints.

\subsubsection{Registration Resolution}
The first phase recovers publication evidence and converts it into publication facts
\(
P=\{\langle t,n_p,h_p,\sigma_p\rangle\},
\)
where each fact records that tool name \(t\) is exposed at syntax node \(n_p\) with candidate handler \(h_p\) under witness \(\sigma_p\). Our implementation models 10 publication variants across Python and JS/TS. \autoref{tab:entrypoint_patterns} summarizes representative publication and dispatch patterns observed in our corpus, including direct declarations, explicit registration APIs, and protocol-level publication patterns. For each matched site, \tool{} resolves aliases, imports, helper wrappers, and tool objects to a normalized tuple \(\langle t,n_p,h_p,\sigma_p\rangle\). When a tool is represented as an object, the collector further resolves handler fields such as \texttt{execute}, \texttt{handler}, or \texttt{callback}, and schema fields such as \texttt{parameters}, \texttt{inputSchema}, or \texttt{schema}. In \autoref{fig:example}, this phase produces a publication fact for \texttt{fetch\_html} even though the concrete callee is revealed only later by the call handler.

\subsubsection{Dispatcher Resolution}
Publication evidence alone is insufficient because many MCP servers route invocations through generic protocol handlers or registries. \tool{} therefore extracts dispatch facts
\(
D=\{\langle t,n_d,h_d,r,\sigma_d\rangle\}
\)
by resolving 16 dispatcher variants, where each fact records that invocations of tool \(t\) are handled by branch \(r\) rooted at node \(n_d\) and forwarded to handler \(h_d\) under witness \(\sigma_d\). These idioms fall into four families. \emph{Protocol dispatch} includes \texttt{setRequestHandler(CallToolRequestSchema, ...)} in JS/TS and \texttt{@call\_tool()} handlers in Python. \emph{Branch dispatch} covers equality tests, membership tests, \texttt{if}/\texttt{elif} chains, \texttt{switch}, and \texttt{match/case} over tool names, as well as outer JSON-RPC routers that first branch on \texttt{tools/call} and then on \texttt{params.name}. \emph{Registry dispatch} covers literal maps, indexed assignments, \texttt{update(...)} extensions, helper-constructed registries, \texttt{map[name]}, \texttt{map.get(name)}, and getter helper indirection. \emph{Reflective dispatch} covers patterns such as \texttt{getattr(self, f"handle\_\{name\}")}. For each recovered dispatcher, \tool{} emits a fact \(\langle t,n_d,h_d,r,\sigma_d\rangle\), where \(r\) records the matched branch scope and \(h_d\) is obtained by resolving forwarded helper calls, registry entries, object methods, or reflective targets. For the example in \autoref{fig:example}, this phase isolates the branch guarded by \texttt{request.params.name == "fetch\_html"} and resolves the forwarded callee to \texttt{Fetcher.html(validatedArgs)}.

\subsubsection{Entrypoint Specialization}
After recovering publication and dispatch facts, \tool{} specializes them into tool-specific analysis units. Let \(P_t\) and \(D_t\) denote the subsets of \(P\) and \(D\) associated with tool name \(t\). For each tool \(t\), \tool{} constructs recovered entrypoints of the form \(e=\langle t,h,r\rangle\), where \(h\) is the resolved handler body and \(r\) is the relevant dispatcher scope. When both publication and dispatch facts are available, \tool{} prefers the handler recovered from dispatch to connect publication with concrete execution. When no dispatch fact is recovered for \(t\), \tool{} falls back to the publication-side handler and sets \(r=\top\) to denote an unbranched publication path. This specialization is important because multiple tools often share a protocol handler, helper layer, or registry object. By materializing \(\langle t,h,r\rangle\), \tool{} seeds taint only along the path reachable for tool \(t\), rather than across the entire dispatcher. Applied to \autoref{fig:example}, the generic request handler is specialized into the concrete \texttt{fetch\_html} entrypoint with handler \texttt{Fetcher.html} and the branch scope for the \texttt{fetch\_html} case.

\subsection{MCP-Specific Taint Specification}
\label{subsec:source-sink}
Once MCP entrypoints are recovered, \tool{} binds taint semantics to MCP-specific trust boundaries to address \emph{Challenge~2}. A repository-wide source--sink policy would be too noisy: not every parameter is user-controlled, and not every return is exposed through the protocol. We therefore define four analysis object classes as shown in \autoref{tab:taint_spec}. \(S_{\mathrm{req}}\) and \(K_{\mathrm{op}}\) support request-side misuse detection, while \(S_{\mathrm{ext}}\) and \(K_{\mathrm{ret}}\) support return-side reasoning about reflected or re-exposed untrusted content.

\begin{table}[htbp]
\centering
\small
\caption{Taint specification instantiated in \autoref{fig:example}.}
\label{tab:taint_spec}
\resizebox{1\linewidth}{!}{
\begin{tabular}{|c|c|}
\hline
\textbf{Category} & \textbf{Running Example Instance} \\
\hline
Request boundary source (\(S_{\mathrm{req}}\)) & \texttt{request.params.arguments} \\ \hline
Sensitive operation sink (\(K_{\mathrm{op}}\)) & \texttt{fetch(url, \{ headers \})} \\ \hline
External content source (\(S_{\mathrm{ext}}\)) & \texttt{response.text()} \\ \hline
Protocol return sink (\(K_{\mathrm{ret}}\)) & \texttt{return \{ content: [\{ text: html \}] \}} \\ \hline

\end{tabular}}
\end{table}

\subsubsection{Request-Side Taint Semantics}
\begin{figure}[t]
    \centering
    \includegraphics[width=1\linewidth]{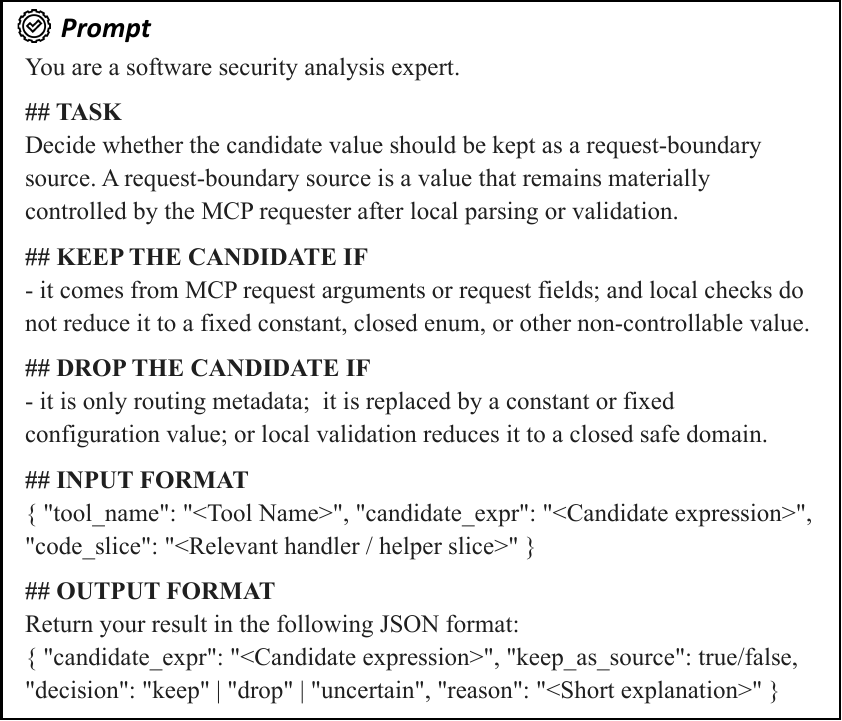}
    \caption{Prompt template for request source identification.}
    \label{fig:prompt}
\end{figure}

For directly exposed MCP tools, formal parameters of the recovered handler are treated as request-boundary sources. For dispatcher-style handlers, taint is introduced only into the non-routing parameters or local variables of the matched branch, which prevents one tool path from polluting another. In addition, \tool{} lifts source semantics through common structured-input accessors, such as \texttt{arguments["path"]}, \texttt{arguments.get("url")}, object destructuring, and schema validated local aliases. This refinement is important because practical MCP handlers often unpack request objects immediately after validation and then propagate only the unpacked locals through deeper helper layers. 
To reduce over-approximation, \tool{} refines these candidates with a lightweight source-controllability check. Straightforward cases are handled deterministically, whereas semantically ambiguous candidates are resolved with LLM assistance only to determine whether the value remains materially requester-controlled after local validation or decoding. A simplified prompt template is shown in \autoref{fig:prompt}.
In \autoref{fig:example}, the request-boundary source is \texttt{request.params.arguments}, even though the relevant \texttt{url} value is recovered only after specific parsing and validation.
Sensitive operation sinks for request-side analysis are modeled through rule-driven specifications over concrete library calls. The current sink packs cover command and code execution, filesystem access, outbound network requests, database and query execution, and related side-effecting operations. Each sink specification binds taint checking to both a callee signature and the argument positions whose taint matters to security. In \autoref{fig:example}, \texttt{fetch(url, \{...\})} is treated as a sensitive operation sink because it issues an outbound request to a target derived from requester-controlled input, even though the code applies a private-IP guard.

\subsubsection{Return-Side Taint Semantics}
Return-side MCP vulnerabilities frequently originate not from client parameters, but from third-party content that the server fetches, reads, or derives from external state before returning it to the host or the model. \tool{} therefore models externally obtained values as a separate source class \(S_{\mathrm{ext}}\). Deterministic source recognizers cover common external-content carriers, including network response bodies, parsed remote JSON, file contents, subprocess \texttt{stdout}/\texttt{stderr}, git logs, markdown or documentation text, pod logs, and similar repository- or runtime-derived artifacts. However, purely syntactic evidence is not always sufficient: some wrappers transform, rename, or partially decode external content before returning it. For such cases, \tool{} slices the producer-side code around the recovered value and invokes an LLM only to classify whether the value is semantically an untrusted external-content carrier. 
In \autoref{fig:example}, \texttt{response.text()} is recognized as an external-content source because it reads the body of a remote HTTP response and directly contributes to the returned HTML.
For return-side analysis, the return value of each recovered MCP handler is treated as a protocol-visible sink \(K_{\mathrm{ret}}\), because it crosses the MCP server boundary and becomes observable to the client, the host, or the model. Unlike a naïve ``every return is a finding'' policy, \tool{} reports a return sink only when the symbolic return value still depends on an externally obtained source or other tainted content. In the running example, the returned \texttt{text: html} field is a return sink because it exposes the remote response body to the MCP client.

\subsection{Bidirectional Interprocedural Taint Analysis}
\label{subsec:propagation}
After entrypoint recovery and taint specification, \tool{} performs bidirectional interprocedural taint analysis over the recovered handlers and helpers. It targets the two trust-boundary flows in our threat model: request-side propagation, in which request-boundary taint reaches a sensitive sink, and return-side propagation, in which externally obtained taint reaches a protocol-visible return. Let \(CG=(F,C)\) denote the call graph over these procedures, where \(F\) is the set of procedures and \(C \subseteq F \times F\) is the call relation. We track taint labels \(\tau(v) \subseteq \{\mathsf{req}, \mathsf{ext}\}\), where \(\mathsf{req}\) denotes request-boundary taint and \(\mathsf{ext}\) denotes externally obtained taint. \autoref{fig:propagation-rules} summarizes the core propagation rules.

\subsubsection{Intra-Procedural Flow Construction}
For each recovered procedure \(p\), \tool{} builds intra-procedural control-flow graph
\(
\mathsf{CFG}_p = (\mathcal{N}_p, \mathcal{E}_p),
\)
where \(\mathcal{N}_p\) is the set of statements in \(p\) and \(\mathcal{E}_p\) is the set of control-flow edges. Over \(\mathsf{CFG}_p\), the analysis derives value-flow relations that capture how tainted data is propagated through assignments, computations, field accesses, container operations, calls, and returns. This step is important because MCP handlers often unpack structured request objects, pass them through helpers, and assemble return objects incrementally rather than flowing taint directly from parameter to sink. At this level, taint continuity is governed by three recurring local patterns: if a tainted value \(y\) is assigned to \(x\), then \(x\) remains tainted (Rule~1); if a tainted component \(v\) is written into a structured location such as \(o.f\), or later read back from it, the taint is preserved across that object boundary (Rule~2); and if a response object \(r\) is assembled from a tainted component \(v\), the enclosing return structure \(r\) is also tainted (Rule~5).

\begin{figure}[t]
\centering
\fbox{
\begin{minipage}{0.95\linewidth}
\footnotesize
\textbf{Notations:}
$\,\alpha \in \{\mathsf{req}, \mathsf{ext}\}$: taint kind;
$\tau(x)=\alpha$: value $x$ carries taint $\alpha$;
$a_i$: actual argument;
$p_i$: formal parameter;
$ret_f$: return value of function $f$;
$r$: assembled return object;
$K$: sensitive sink.

\textbf{Rules:}
\begin{enumerate}[leftmargin=12pt, itemsep=2pt, label=\arabic*.]
    \item \textbf{Assignment Propagation:}
    $\tau(y)=\alpha \wedge \sigma: x := y \Longrightarrow \tau(x)=\alpha$
    \item \textbf{Field / Container Propagation:}
    $\tau(v)=\alpha \wedge \sigma: o.f := v \Longrightarrow \tau(o.f)=\alpha$\\
    $\tau(o.f)=\alpha \wedge \sigma: x := o.f \Longrightarrow \tau(x)=\alpha$
    \item \textbf{Call-Argument Propagation:}
    $\tau(a_i)=\alpha \wedge \sigma: call~f(a_1,\ldots,a_n) \Longrightarrow \tau(p_i)=\alpha$
    \item \textbf{Return Propagation:}
    $\tau(ret_f)=\alpha \wedge \sigma: x := f(\ldots) \Longrightarrow \tau(x)=\alpha$
    \item \textbf{Return-Assembly Propagation:}
    $\tau(v)=\alpha \wedge \sigma: r := \{\ldots, v, \ldots\} \Longrightarrow \tau(r)=\alpha$
    \item \textbf{Sink Reachability:}
    $\tau(v)=\alpha \wedge \sigma: K(v) \Longrightarrow$ emit candidate path
\end{enumerate}
\end{minipage}
}
\caption{Core taint propagation rules in \tool{}.}
\label{fig:propagation-rules}
\end{figure}

\subsubsection{Inter-Procedural Forward Propagation}
Forward propagation starts from \(S_{\mathrm{req}}\) and follows interprocedural value dependencies toward \(K_{\mathrm{op}}\). Across call-graph edges, \tool{} propagates taint in two complementary steps summarized by Rules~3--4 of \autoref{fig:propagation-rules}: a tainted actual argument \(a_i\) taints the corresponding callee parameter \(p_i\), and a tainted return value \(ret_f\) taints the receiving caller-side variable \(x\). This process computes the request-reachable region
\[
\mathcal{R}^{f}_{\mathrm{op}} = \mathsf{ReachForward}(S_{\mathrm{req}}, CG),
\]
which conservatively over-approximates all procedures and symbolic values that may be influenced by requester-controlled data. In \autoref{fig:example}, taint originates at \texttt{request.params.arguments}, is preserved in \texttt{validatedArgs.url} after schema parsing, and then propagates through \texttt{Fetcher.html} and \texttt{\_fetch} until it reaches the \texttt{fetch(url, ...)} operation sink. The private-IP check is treated as a guard on the path rather than as a transformation that removes request-boundary taint.

\subsubsection{Inter-Procedural Backward Propagation}
Backward propagation starts from \(K_{\mathrm{ret}}\) and traces the symbolic contributors of each protocol-visible return value back toward candidate producers. The goal is to determine whether a returned value depends on an element of \(S_{\mathrm{ext}}\), rather than on a fixed literal, a local wrapper string, or a self-constructed status object. Operationally, this phase reuses the same summary relations in reverse: from a returned caller value \(x\), Rule~4 steps back into the callee return \(ret_f\), while Rule~5 peels apart an assembled response object \(r\) to recover the concrete producer values \(v\) that contribute to the protocol-visible field. Formally, this phase computes the return-reachable region
\[
\mathcal{R}^{b}_{\mathrm{ret}} = \mathsf{ReachBackward}(K_{\mathrm{ret}}, CG),
\]
which includes assignments, helper returns, content-assembly statements, and producer-side I/O operations that contribute to the returned object. In the running example of \autoref{fig:example}, the returned \texttt{html} field is traced backward through \texttt{applyLengthLimits(html, ...)} to \texttt{response.text()}, which is already classified as an external-content source. This backward procedure is especially important for MCP because response-side risks are rarely visible as direct source-to-sink pairs; instead, the analysis must reconstruct how data is obtained, transformed, and eventually serialized into a client-visible return object.

\subsubsection{Intersection and Path Refinement}
The final phase validates complete taint paths by intersecting source-reachable and sink-reachable regions. Candidate findings are emitted only when the path construction induced by the earlier propagation steps still terminates at a sensitive sink, i.e., when some symbolic value \(v\) remains tainted at a sink invocation \(K(v)\); this final reachability condition is captured by Rule~6 in \autoref{fig:propagation-rules}. For operation flows, \tool{} computes
\[
\mathcal{I}_{\mathrm{op}} = \mathcal{R}^{f}(S_{\mathrm{req}}) \cap \mathcal{R}^{b}(K_{\mathrm{op}}),
\]
and retains a candidate only when the intersection forms a valid intra- or interprocedural data-flow path from a request-boundary source to a sensitive operation sink. For return flows, it computes
\[
\mathcal{I}_{\mathrm{ret}} = \mathcal{R}^{f}(S_{\mathrm{ext}}) \cap \mathcal{R}^{b}(K_{\mathrm{ret}}),
\]
and retains a candidate only if the returned symbolic value still depends on an external-content source. After intersection, \tool{} performs path refinement to suppress candidates invalidated by effective guards or sanitizers, such as strict allowlists, canonicalization followed by root-membership checks for file paths, parameterized query APIs, or fixed-base URL constructions. The analysis records guard evidence, including schema checks, confinement checks, and quoting or normalization routines, along each candidate path. Straightforward cases are handled by deterministic suppression rules, whereas only semantically ambiguous guards are deferred to LLM-based adjudication to determine whether attacker influence is materially eliminated. Finally, equivalent paths that correspond to the same recovered MCP handler, sink site, and root cause are merged, and one representative finding is retained for downstream review.

\section{Evaluation}

Our evaluation focuses on three aspects of \tool{}: detection effectiveness on confirmed cases, applicability to real-world MCP servers, and the role of its major components.

\noindent\hangindent=2.5em\hangafter=1\textbf{RQ1: How well does \tool{} detect confirmed MCP vulnerabilities relative to existing tools?}
We study this question on a benchmark of confirmed MCP vulnerability cases, using case-level detections and recall for comparison.

\noindent\hangindent=2.5em\hangafter=1\textbf{RQ2: What vulnerabilities can \tool{} uncover in real-world MCP servers?}
To answer this question, we apply \tool{} to real-world open-source MCP server repositories and manually review the resulting candidate findings.

\noindent\hangindent=2.5em\hangafter=1\textbf{RQ3: Which components are responsible for \tool{}'s coverage and triage behavior?}
An ablation study is used to isolate how the major MCP-aware components affect reviewed-instance coverage and triage outcomes.

\begin{table*}[t]
\centering
\fontsize{7.2}{8.8}\selectfont
\caption{Case-level comparison on the 32 confirmed MCP vulnerability cases.}
\label{tab:baseline}

\renewcommand{\arraystretch}{1.10}
\setlength{\tabcolsep}{3.8pt}

\begin{tabular}{|>{\centering\arraybackslash}m{1.9cm}|>{\raggedright\arraybackslash}m{1.8cm}|c|c|c|c|c|c|c|c|c|}
\hline
\centering\textbf{Type} & \centering\textbf{CVE ID} & \centering\textbf{Server Name} & \textbf{Lang.} & \textbf{Req.} & \textbf{Resp.} & \textbf{CodeQL} & \textbf{Semgrep} & \textbf{Snyk Code} & \textbf{MCPScan} & \textbf{\tool} \\
\hline

\multirow{19}{*}{\makecell[c]{\textbf{Command /}\\\textbf{Code Execution}}}
& CVE-2025-11202 & win-cli-mcp-server & JS/TS & \cmark &        & \no  & \yes & \no  & \yes & \yes \\
& CVE-2025-52573 & ios-simulator-mcp & JS/TS & \cmark &        & \no  & \no  & \yes & \yes & \yes \\
& CVE-2025-5277  & aws-mcp-server & Py    & \cmark &        & \no  & \no  & \no  & \yes & \yes \\
& CVE-2025-53818 & github-kanban-mcp-server & JS/TS & \cmark &        & \no  & \no  & \no  & \no  & \yes \\
& CVE-2025-53832 & lara-mcp & JS/TS & \cmark &        & \no  & \no  & \yes & \no  & \yes \\
& CVE-2025-59834 & adb-mcp & JS/TS & \cmark &        & \no  & \no  & \no  & \no  & \yes \\
& CVE-2025-63603 & mcp-server-data-exploration & Py    & \cmark &        & \no  & \yes & \no  & \yes & \yes \\
& CVE-2025-69256 & serverless & JS/TS & \cmark &        & \no  & \no  & \no  & \no  & \yes \\
& CVE-2026-26029 & sf-mcp-server & JS/TS & \cmark &        & \no  & \no  & \no  & \yes & \yes \\
& CVE-2025-53107 & git-mcp-server & JS/TS & \cmark & \cmark & \no  & \no  & \no  & \no  & \yes \\
& CVE-2025-53355 & mcp-server-kubernetes & JS/TS & \cmark & \cmark & \no  & \no  & \no  & \yes & \yes \\
& CVE-2025-66404 & mcp-server-kubernetes & JS/TS & \cmark & \cmark & \no  & \no  & \no  & \yes & \yes \\
& CVE-2025-58358 & markdownify-mcp & JS/TS & \cmark & \cmark & \yes & \no  & \yes & \yes & \yes \\
& CVE-2025-54073 & mcp-package-docs & JS/TS & \cmark & \cmark & \no  & \no  & \yes & \no  & \yes \\
& CVE-2025-53967 & Figma-Context-MCP & JS/TS & \cmark & \cmark & \no  & \no  & \no  & \no  & \yes \\
& CVE-2025-53372 & node-code-sandbox-mcp & JS/TS & \cmark & \cmark & \no  & \no  & \no  & \yes & \yes \\
& CVE-2025-59376 & mcp-kubernetes-server & Py    & \cmark &        & \no  & \no  & \no  & \yes & \yes \\
& CVE-2025-59377 & mcp-kubernetes-server & Py    & \cmark &        & \no  & \yes & \no  & \yes & \yes \\
& CVE-2026-2178  & xcode-mcp-server & JS/TS & \cmark &        & \no  & \no  & \no  & \no  & \no \\
\hline

\multirow{10}{*}{\centering\makecell[c]{\textbf{Filesystem}\\\textbf{Access}}}
& CVE-2025-53109 & server-filesystem & JS/TS & \cmark &        & \no & \yes & \yes & \no & \yes \\
& CVE-2025-53110 & server-filesystem & JS/TS & \cmark &        & \no & \yes & \yes & \no & \yes \\
& CVE-2025-67364 & fast-filesystem-mcp & JS/TS & \cmark &        & \no & \yes & \yes & \no & \yes \\
& CVE-2025-68143 & mcp-server-git & Py    & \cmark &        & \no & \no  & \no  & -   & \yes \\
& CVE-2025-68144 & mcp-server-git & Py    & \cmark &        & \no & \no  & \no  & -   & \yes \\
& CVE-2025-68145 & mcp-server-git & Py    & \cmark &        & \no & \no  & \no  & -   & \yes \\
& CVE-2026-27735 & mcp-server-git & Py    & \cmark &        & \no & \no  & \no  & -   & \yes \\
& CVE-2026-27825 & mcp-atlassian & Py    & \cmark &        & \no & \no  & \no  & -   & \yes \\
& CVE-2025-61685 & mastra & JS/TS & \cmark &        & \no & \no  & \no  & \no & \no \\
& CVE-2025-5273  & markdownify-mcp & JS/TS & \cmark &        & \no & \yes & \yes & \no & \yes \\
\hline

\makecell[c]{\textbf{Database /}\\\textbf{Query Execution}}
& CVE-2025-59333 & mcp-database-server & JS/TS & \cmark &        & \no & \yes & \no & - & \yes \\
\hline

\multirow{2}{*}{\centering\textbf{SSRF}}
& CVE-2025-5276  & markdownify-mcp & JS/TS & \cmark &        & \no & \no & \yes & \no & \yes \\
& CVE-2025-65513 & fetch-mcp & JS/TS & \cmark & \cmark & \no & \no & \yes & \no & \yes \\
\hline

\multicolumn{6}{|c|}{\textbf{Total Detected (32 Cases)}}
& \textbf{1} & \textbf{8} & \textbf{10} & \textbf{11} & \textbf{30} \\
\hline
\end{tabular}

\vspace{2pt}
\begin{minipage}{0.98\textwidth}
\footnotesize
\textbf{Note.} \textbf{Lang.}: \texttt{Py} = Python and \texttt{JS/TS} = JavaScript/TypeScript. In the \textbf{Req.} and \textbf{Resp.} columns, \cmark\ indicates that the case is included in the corresponding request-side or return-side split; rows marked in both columns denote mixed chains. In the tool columns, \yes\ denotes a reviewed case-compatible detection and \no\ denotes no credited detection. A dash (`-') indicates that the case lies outside the evaluated capability boundary of that tool. 
\end{minipage}
\end{table*}

\subsection{Evaluation Setup}
\label{subsec:setup}

\noindent\textbf{Tool implementation.}
We implemented a prototype of \tool{} on top of the open-source static analysis framework \texttt{YASA}~\cite{yasa}. The pipeline consists of three main stages: MCP entrypoint recovery, MCP-specific taint specification, and bidirectional interprocedural taint analysis. Semantically ambiguous source and guard cases are handled through the LLM-assisted adjudication.

\noindent\textbf{Environment.}
All experiments were conducted on a server running Ubuntu 22.04, equipped with two 64-core AMD EPYC 9954 processors, 1024\,GB of RAM, and six NVIDIA A100 GPUs, each with 80\,GB of memory. The LLM-assisted analysis in our pipeline was powered by the gpt-5.3-codex-medium API.

\noindent\textbf{Baselines.}
We compare \tool{} against four baseline tools: \texttt{CodeQL}~\cite{codeql}, \texttt{Semgrep}~\cite{semgrep}, \texttt{Snyk Code}~\cite{snykcode}, and \texttt{MCPScan}~\cite{mcpscan}. \texttt{CodeQL} serves as a representative query-based static analysis baseline, while \texttt{Semgrep} represents a lightweight rule-based analyzer. \texttt{Snyk Code} is a commercial semantic SAST baseline with announced support for MCP-specific input sources, mainly in \texttt{FastMCP}-based implementations. \texttt{MCPScan} is the closest MCP-oriented baseline in our comparison, combining rule matching with LLM-assisted triage.

\noindent\textbf{Dataset.}
For RQ1, we collected MCP-related vulnerability cases from public sources including GitHub Security Advisories, the GitLab Advisory Database, NVD, OSV, and selected security reports. Cases were retained through manual curation requiring clear MCP relevance, convincing security evidence, and implementations in Python, JavaScript, or TypeScript. The finalized comparative benchmark contains 32 confirmed vulnerability cases spanning several categories, with both request-side and return-side relevant scenarios represented.
For RQ2, we gathered MCP server repositories from three public registries: \textit{mcp.so}, \textit{PulseMCP}, and \textit{MCPWorld}. Starting from 75,380 GitHub repository links, URL normalization and cross-source deduplication produced 63,639 unique repositories. We then enriched each repository with GitHub metadata and applied rule-based filtering to retain accessible repositories likely to implement MCP servers, removing clients, SDKs, examples, templates, registries, and archived or forked projects. Our analysis concentrates on Python, JavaScript, and TypeScript because these three languages collectively account for nearly 80\% of the collected MCP servers, making them the most representative target for large-scale study. The corpus contains 15,452 repositories: 8,047 in Python (52.1\%), 4,977 in TypeScript (32.2\%), and 2,428 in JavaScript (15.7\%).

\subsection{RQ1: Detection Effectiveness}

To answer RQ1, we evaluate \tool{} on the MCP vulnerability benchmark of 32 confirmed cases drawn from Python, JavaScript, and TypeScript MCP servers, covering command and code execution, filesystem access, database and query execution, and SSRF. Because the benchmark contains only vulnerable cases, we report case-level detections and recall rather than precision. As shown in \autoref{tab:baseline}, \tool{} detects \textbf{30 of 32 cases} (93.8\% recall).

The benchmark is dominated by request-side cases and mixed chains rather than pure return-only disclosures, since public MCP advisories are typically anchored in a concrete side-effecting sink. Rows marked in both the \emph{Req.} and \emph{Resp.} columns should be read as \emph{mixed chains}: attacker-relevant content crosses the MCP return boundary and later contributes to a downstream request-side effect. Against this benchmark, \texttt{CodeQL} detects only 1 case, as its default queries do not model MCP-specific tool exposure, handler parameters, or protocol-visible returns as analysis boundaries. \texttt{Semgrep} reaches 8 cases when the bug manifests as a local API misuse, but loses coverage once the flow spans helpers, wrappers, or indirect dispatch. \texttt{Snyk Code} detects 10 cases, partly because it already recognizes MCP server sources and entrypoints in FastMCP and the official Python and TypeScript SDKs, yet it still misses vulnerabilities that require broader dispatch recovery, wrapper traversal, or return-side propagation. \texttt{MCPScan} is the strongest baseline at 11 cases, though it falls well short of \tool{}, showing that MCP-specific rules are insufficient without end-to-end recovery of both propagation directions.

The two false negatives mark the current implementation boundary in JavaScript. CVE-2025-65513 requires rebinding decoded MCP arguments into object-destructured parameters with default values, while CVE-2025-61685 requires propagating taint into callback parameters across a higher-order iteration and closure boundary. Both misses trace back to current gaps in parameter rebinding and cross-closure taint propagation.

\subsection{RQ2: Real-World Applicability}

To answer RQ2, we ran \tool{} on 15,452 MCP server repositories and manually reviewed the highest-risk outputs. Near-duplicate traces from the same server and nearby sink site were merged into \emph{overlap-compressed candidate clusters}, yielding 549 clusters across 424 servers. Disclosure review then consolidated surviving evidence into distinct \emph{vulnerability paths}, confirming 118 paths across 87 servers.
In the first review pass, 278 clusters were marked reportable, 55 required additional manual judgment, and 216 were rejected (details are included in our artifact). Rejected clusters fell into a few recurring patterns: user input did not actually control the request target in 122 cases, the target was effectively fixed or configuration-only in 53, and local validation made shell interpolation non-exploitable in 21. The final disclosure review applied stricter criteria, confirming 118 paths and rejecting the remaining 431 cases. The gap between candidates and confirmed paths reflects recall-oriented candidate generation paired with conservative disclosure standards rather than a direct precision estimate.
Among the 118 confirmed paths, 111 involve code execution, 5 involve query execution, and 2 involve SSRF. The primary sink functions are summarized in \autoref{tab:sink_rules}.

\begin{table}[t]
\centering
\small
\caption{Primary sink functions in the confirmed real-world MCP vulnerabilities (\(N=118\)).}
\label{tab:sink_rules}
\resizebox{0.97\linewidth}{!}{
\begin{tabular}{|c|c|c|}
\hline
\textbf{Vulnerability Type} & \textbf{Sink Function / Primitive} & \textbf{\# Paths} \\
\hline

\multirow{8}{*}{\textbf{Code Execution}}
& \texttt{execAsync} & 60 \\
& \texttt{execSync} & 28 \\
& \texttt{exec} & 9 \\
& \texttt{eval} & 5 \\
& \texttt{Function constructor} & 3 \\
& \texttt{conn.exec} & 2 \\
& \texttt{asyncio.create\_subprocess\_shell} & 2 \\
& \texttt{subprocess.run} & 2 \\
\hline

\multirow{3}{*}{\textbf{Query Execution}}
& \texttt{connection.query} & 3 \\
& \texttt{client.query} & 1 \\
& \texttt{db.unsafe} & 1 \\
\hline

\textbf{SSRF}
& \texttt{fetch} & 2 \\

\hline
\end{tabular}}
\vspace{2pt}
\begin{minipage}{0.97\linewidth}
\end{minipage}
\end{table}

We next present two representative case studies corresponding to the two propagation directions in our threat model: a request-side path from MCP tool input to a privileged execution sink, and a return-side path in which externally obtained content crosses trust boundaries and contributes to unsafe downstream behavior.

\begin{figure}[t!]
    \centering
    \includegraphics[width=0.96\linewidth]{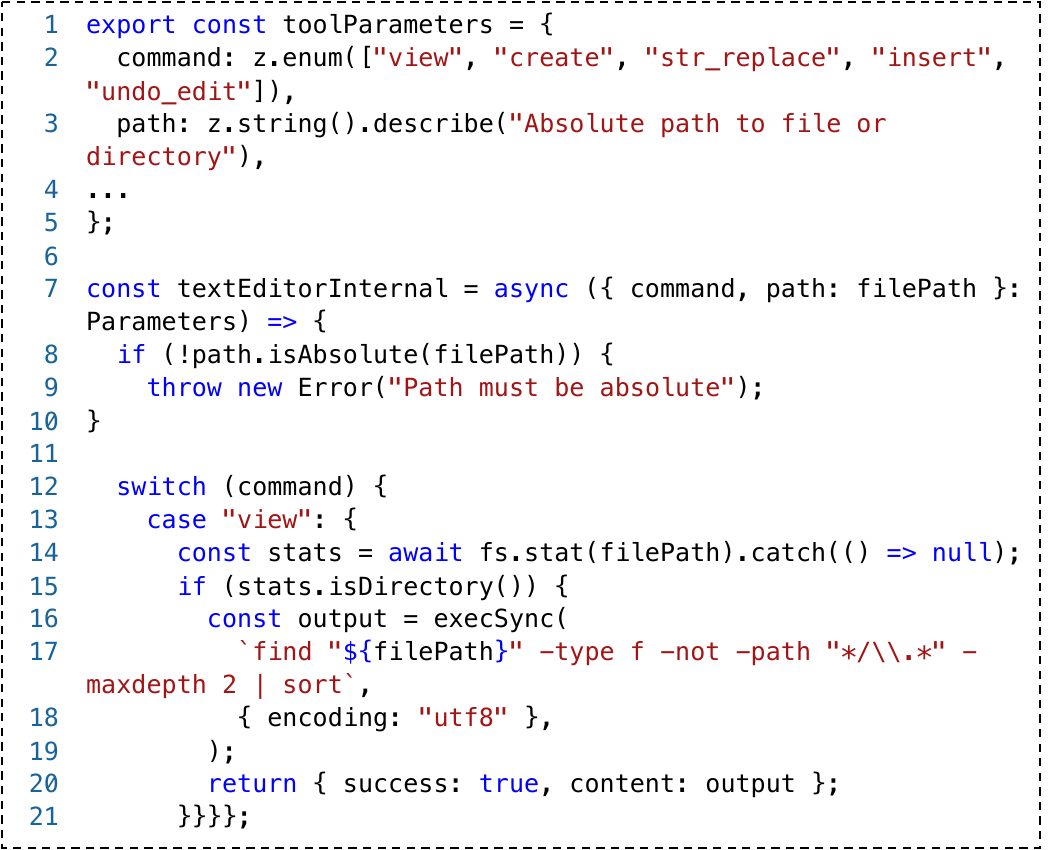}
    \caption{Request-side propagation to command execution in \texttt{mcp-server-text-editor}.}
    \label{fig:case1}
\end{figure}

\noindent\textbf{Case study \#1: Request-side propagation.}
Our first case comes from \texttt{bhouston/mcp-server-text-editor}, an MCP implementation of Claude's built-in text editor. Although the tool is presented as a file-viewing and editing interface, the \texttt{view} path invokes \texttt{execSync} when the supplied \texttt{path} refers to a directory. As shown in \autoref{fig:case1}, the \texttt{path} parameter is requester-controlled and verified only to be absolute before being interpolated into a shell command. That check offers no protection against shell injection, since an absolute path can still carry shell-significant characters. The resulting flow is a representative request-side propagation path: MCP-visible input crosses the server boundary and reaches a privileged execution sink, allowing a seemingly benign tool to be steered into arbitrary shell execution on the host.

\begin{figure}[t!]
    \centering
    \includegraphics[width=0.96\linewidth]{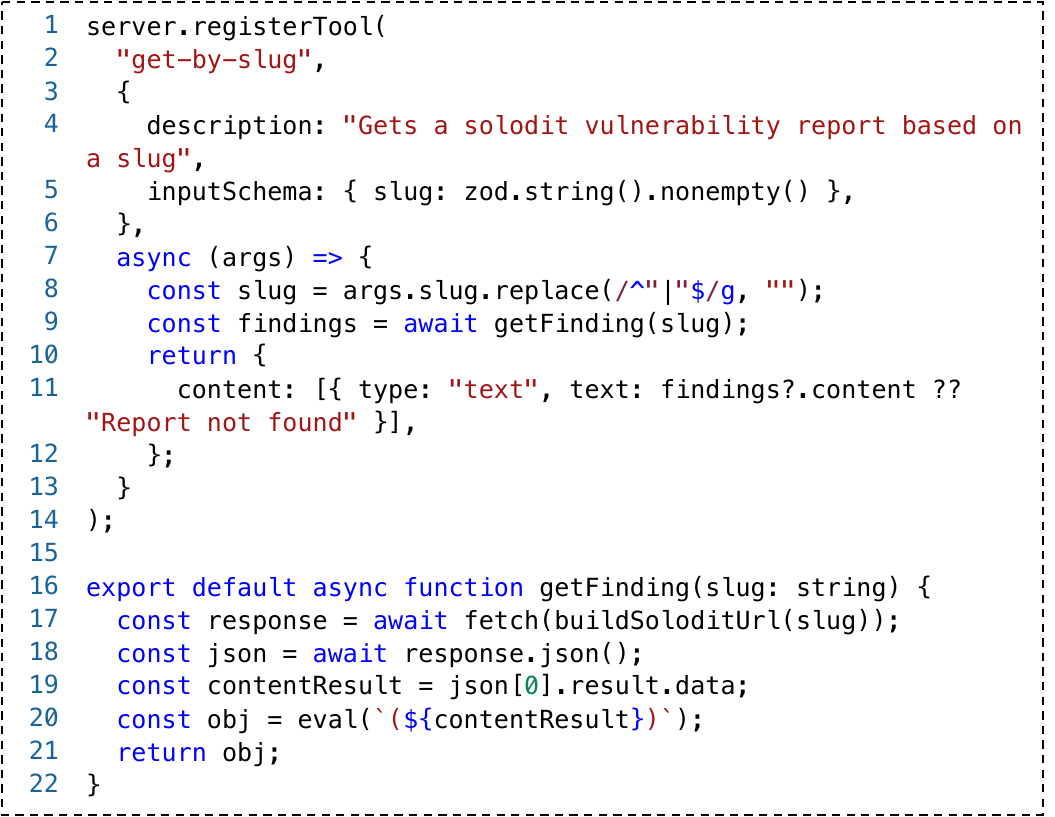}
    \caption{Externally sourced propagation with return-side risk in \texttt{search-solodit-mcp}.}
    \label{fig:case2}
\end{figure}

\noindent\textbf{Case study \#2: Externally sourced propagation and return-side risk.}
\texttt{LyuboslavLyubenov/search-solodit-mcp} is our second case, a server that searches and retrieves public Solodit vulnerability reports. Here, the security-relevant data does not originate from a local tool argument flowing directly into a sink. Instead, the \texttt{get-by-slug} tool accepts a user-controlled \texttt{slug}, fetches report data from the Solodit API, extracts \texttt{json[0].result.data}, and passes it directly to \texttt{eval}, as shown in \autoref{fig:case2}, with no structured decoding or validation applied before evaluation. This case illustrates that unsafe propagation in MCP servers is not limited to direct request-side misuse: externally obtained content can enter through a retrieval path, remain security-relevant across the MCP interaction, and reach a dangerous operation several steps later. In our threat model, such flows connect closely to return-side risk, because the same externally sourced content is delivered through an MCP-visible retrieval interface and may participate in mixed chains that influence later model reasoning or tool invocations.

\subsection{RQ3: Ablation Study}

We answer RQ3 through a variant-based ablation on the reviewed real-world instance set. \autoref{tab:rq3_ablation_projects} compares the full pipeline against three ablated variants, each omitting one core component: M1 for MCP entrypoint recovery, M2 for LLM-assisted source identification, and M3 for LLM-assisted guard and sanitizer validation. Results are reported over reviewed instances rather than unique servers. The reviewed set contains 859 instances in total, of which 424 are confirmed, 302 unresolved, and 133 rejected. These counts are not directly comparable to the server- and cluster-level figures in RQ2, since a single server may contribute multiple reviewed instances with different paths, sinks, or review outcomes.
M1 and M2 both bear primarily on coverage. Removing M1 reduces surfaced instances from 859 to 636 and confirmed instances from 424 to 287, which reflects how much MCP-specific entrypoint recovery contributes to exposing handlers hidden behind protocol dispatch and wrapper code. Removing M2 leaves 621 surfaced instances and 324 confirmed ones, indicating that many findings depend on recovering source semantics beyond direct tool parameters.
M3 plays a distinct role. Its removal leaves the confirmed count unchanged at 424, but unresolved instances rise from 302 to 368 and rejected instances fall from 133 to 67. The component therefore improves triage quality rather than raw coverage: it converts a substantial fraction of otherwise ambiguous cases into defensible reject decisions, bringing the retained set down from 792 to 726. Taken together, the ablation indicates that M1 and M2 govern finding coverage while M3 governs the reliability of triage outcomes.

\begin{table}[t]
\centering
\small
\caption{Ablation on reviewed real-world instances.}
\label{tab:rq3_ablation_projects}

\renewcommand{\arraystretch}{1.08}
\setlength{\tabcolsep}{4.8pt}
\resizebox{1\linewidth}{!}{
\begin{tabular}{lrrrrr}
\hline
\textbf{Variant} & \textbf{Confirmed} & \textbf{Unresolved} & \textbf{Rejected} & \textbf{Retained} & \textbf{Total} \\
\hline
M1+M2+M3 & 424 & 302 & 133 & 726 & 859 \\
w/o M1 & 287 & 251 & 98 & 538 & 636 \\
w/o M2 & 324 & 210 & 87 & 534 & 621 \\
w/o M3 & 424 & 368 & 67 & 792 & 859 \\
\hline
\end{tabular}}

\vspace{2pt}
\begin{minipage}{0.97\linewidth}
\footnotesize
\textbf{Note.} The reviewed set contains 859 reviewed instances: 424 confirmed, 302 unresolved, and 133 rejected. These are reviewed instances rather than unique servers; one server may contribute multiple instances. M1: MCP entrypoint recovery. M2: LLM-assisted source identification. M3: LLM-assisted guard/sanitizer validation. \textbf{Retained} = Confirmed + Unresolved.
\end{minipage}
\end{table}

\section{Discussion}

\noindent\textbf{Mitigation.}
Our findings suggest that securing MCP servers requires attention to both propagation directions. On the request side, requester-controlled values should be prevented from reaching execution boundaries through shell commands, unconstrained paths, URLs, or query strings. On the return side, externally obtained or sensitive internal data should not be treated as benign tool output by default, since it may later re-enter prompts or downstream tool invocations. Structured argument validation, explicit allowlisting, and deployment safeguards such as sandboxing, filesystem restrictions, and network egress control together reduce the attack surface that MCP tool interfaces expose.

\noindent\textbf{Generality.}
The core design of \tool{} is not tied to a specific MCP SDK or language. It is organized around three abstractions: entrypoint recovery, trust-boundary-specific source and sink modeling, and bidirectional interprocedural propagation. The current implementation targets Python, JavaScript, and TypeScript MCP servers, but the same formulation can accommodate other languages by supplying language-specific support for registration, dispatch, and sink patterns. More broadly, the approach is applicable to other tool-serving frameworks that connect LLM-powered applications to privileged capabilities.

\noindent\textbf{Limitations.}
\tool{} targets unsafe cross-boundary data flows rather than the full space of MCP security problems. Authentication, authorization, business-logic, and deployment flaws fall outside its scope unless they manifest as analyzable data-flow violations, and metadata-level attacks such as tool-description poisoning are not directly addressed unless those artifacts propagate into executable code paths captured by the analysis. On the practical side, false positives remain the main challenge: the pipeline may over-approximate attacker control, overlook effective local constraints, or flag flows whose targets are fixed, configuration-only, or otherwise not meaningfully attacker-influenced. Completeness can also suffer when servers rely on reflective dispatch, runtime-generated handlers, or framework-specific wrappers. Finally, the real-world evaluation is disclosure-oriented, and semantically ambiguous cases require LLM-assisted adjudication.

\section{Related Work}

\noindent\textbf{Research on MCP security.}
As MCP becomes a foundational interface for tool-augmented AI ecosystems, its security and reliability have drawn increasing attention~\cite{hou2025mcp,hasan2025mcpglance}. Most prior work characterizes the MCP attack surface rather than statically recovering vulnerable data flows in server implementations. Hasan et al.~\cite{hasan2025mcpglance} measure open-source MCP servers and identify common vulnerability categories and maintenance weaknesses, while Hou et al.~\cite{hou2025mcp} survey the broader MCP landscape, threat model, and open research directions. Zhao et al.~\cite{zhao2025mcpattack} present a taxonomy of malicious MCP server behaviors and demonstrate proof-of-concept exploits, and Radosevich and Halloran~\cite{radosevich2025mcpsafety} show that current MCP workflows remain vulnerable to severe code-execution and privilege-escalation attacks. Other work examines preference manipulation in open MCP marketplaces~\cite{wang2025mpma}, parasitic toolchain attacks across MCP components~\cite{zhao2025mindyourserver}, and benchmark-driven security evaluation through MCPSecBench~\cite{yang2025mcpsecbench}. Complementary defenses explore OAuth-enhanced tool definitions, protocol hardening, secure gateways, and layered detection mechanisms~\cite{bhatt2025etdi,xing2025mcpguard,kumar2025mcguardian,brett2025mcpgateway}. We take a different angle, using static analysis to recover request-side and return-side trust-boundary violations directly in MCP server code.

\noindent\textbf{Security of tool-using LLM agents.}
Beyond MCP, a growing body of work studies the security risks inherent in tool-using LLM systems and autonomous agents~\cite{li2024personal,he2024securityaiagents,10.1145/3716628,wang2025largemodelbasedagents,yu2025survey}. Tool use expands the attack surface of LLM applications by coupling model reasoning with external actions and untrusted content channels~\cite{he2024securityaiagents,10.1145/3716628,zhang2024agent}. Indirect prompt injection is a particularly well-studied instance of this risk: attacker-controlled content retrieved from external sources can re-enter the model's context and shape subsequent reasoning or tool invocations~\cite{greshake2023not,liu2024demystifying}. This literature connects closely to our return-side threat model, yet it typically analyzes system-level attacks, benchmarks, host-side failures, or agent-level taxonomies without performing static reasoning over MCP server code~\cite{zhang2024agent,yu2025survey}. We make these risks analyzable at the server level by treating protocol-visible returns as first-class security boundaries and tracing the paths through which external content reaches them.

\section{Conclusion}
\label{sec:conclusion}
As the MCP take on a growing role in connecting LLM agents to external tools, the trust-boundary crossings they introduce create classes of unsafe data flow that conventional analyzers are poorly equipped to detect. We framed this as a bidirectional problem and built \tool{} around MCP entrypoint recovery, protocol-specific taint modeling, and interprocedural analysis, covering both request-side and return-side propagation within a unified framework. On a benchmark of 32 confirmed vulnerability cases, \tool{} detects 30, outperforming CodeQL, Semgrep, Snyk Code, and MCPScan. Across 15,452 real-world repositories, it uncovers 118 vulnerability paths in 87 servers, confirming that insecure data flow is not an edge case in the MCP ecosystem but a recurring structural weakness.



\bibliographystyle{ACM-Reference-Format}
\bibliography{acmart}

\end{document}